\documentstyle[preprint,aps,epsfig]{revtex}

\begin{document}
\draft
\title {Optical nonlinearity enhancement of graded metallic films}
\date {\today}
\author {J. P. Huang$^{1,2}$ and K. W. Yu$^1$}
\address {$^1$Department of Physics, The Chinese University of Hong Kong, 
 Shatin, NT, Hong Kong \\
 $^2$Max Planck Institute for Polymer Research, Ackermannweg 10, 
 55128 Mainz, Germany}

\maketitle

\begin{abstract}

The effective linear and third-order nonlinear susceptibility of graded 
metallic films with weak nonlinearity have been investigated.
Due to the simple geometry, we were able to derive exactly the local 
field inside the graded structures having a Drude dielectric gradation 
profile. We calculated the effective linear dielectric constant and 
third-order nonlinear susceptibility.
We investigated the surface plasmon resonant effect on the optical
absorption, optical nonlinearity enhancement, and figure of merit of
graded metallic films. It is found that the presence of gradation in 
metallic films yields a broad resonant plasmon band in the optical region,
resulting in a large enhancement of the optical nonlinearity and hence
a large figure of merit.
We suggest experiments be done to check our theoretical predictions,
because graded metallic films can be fabricated more easily than graded 
particles.

\vskip 5mm
\end{abstract}

%\section{Introduction}

Thin films can possess different optical properties (see, e.g., 
Ref.~\cite{Kammler}) when comparing with their bulk counterparts. 
Also, graded materials have quite different physical properties from the 
homogeneous materials, hence making the composite media consisting of
graded inclusions more useful and interesting. Recently, some authors 
found experimentally that the graded thin films may have better dielectric 
properties than a single-layer film~\cite{LuAPL03}. 
In fact, graded materials~\cite{Milton} are the materials whose material 
properties can vary continuously in space. These materials have attracted
much interest in various engineering applications~\cite{Yamanouchi}.
However, the traditional theories~\cite{Jackson,AIP} fail to deal with the
composites of graded inclusions. Recently, for treating these 
composites, we presented a first-principles approach~\cite{Dong,Gu-JAP} 
and a differential effective dipole approximation~\cite{Yu1,Huang}.

The problem becomes more complicated by the presence of nonlinearity in 
realistic composites. Besides gradation (inhomogeneity), nonlinearity 
plays also an important role in the effective material properties of 
composite 
media~\cite{Stroud,Agarwal,Zeng,DJBergmanPRB89,Scaife,Bergman,YuPRB93,Shalaev,Hui,Sarychev,Gao1}.
There is a great need for nonlinear optical materials with large 
nonlinear susceptibility or optimal figure of merit (FOM). For 
these purposes, many studies have been devoted to achieving a large 
nonlinearity enhancement or optimal FOM of bulk composites by the 
surface-plasmon resonance in metal-dielectric 
composites~\cite{Shalaev,Sarychev}, as well as by taking 
into account the structural information like random distribution of 
particles~\cite{Sipe} and particle shape distribution~\cite{Gao3}. 
More recently, a large nonlinearity enhancement has been found when the 
authors studied a sub-wavelength multilayer of titanium dioxide and 
conjugated polymer~\cite{Fischer}.

Regarding the surface plasmon resonant nonlinearity enhancement, there 
is always a concomitantly strong absorption, that renders the FOM 
of the resonant enhancement peak to be too small to be useful. 
In this work, we will consider graded metallic films to circumvent the 
problem. We have found that the presence of gradation in metallic
films yields a broad resonant plasmon band in the optical region.
We have succeeded in gaining a large nonlinearity enhancement as well as 
optimal FOM by considering the effect of inhomogeneity due to gradation.

%\section{Formalism}

Let us consider a graded metallic film with width $L$, and the gradation under consideration is in the direction perpendicular to the film.  The local 
constitutive relation between the displacement (${\bf D}$) and the 
electric field (${\bf E}$) inside the graded layered geometry is given by
\begin{equation}
{\bf D}(z,\omega) = \epsilon(z,\omega){\bf E}(z,\omega)+\chi(z,\omega)
 |{\bf E}(z,\omega)|^2{\bf E}(z,\omega),
\label{Weak} 
\end{equation}%1
where $\epsilon (z,\omega)$ and $\chi(z,\omega)$ are respectively the 
linear dielectric constant and third-order nonlinear susceptibility. 
Note that both $\epsilon(z,\omega)$ and $\chi(z,\omega)$ are gradation 
profiles as a function of position $r$. 
Here we assume that the weak nonlinearity condition is satisfied, 
that is, the contribution of the second term (nonlinear part 
$\chi(z,\omega)|{\bf E}(z,\omega)|^2$) in the right-hand side of 
Eq.~(\ref{Weak}) is much less than that of the first term 
(linear part $\epsilon(z,\omega)$)~\cite{Stroud}. 
We further restrict our discussion to the quasi-static approximation, 
under which the whole layered geometry can be regarded as an effective 
homogeneous one with effective (overall) linear dielectric constant 
$\bar{\epsilon}(\omega)$ and effective (overall) third-order nonlinear 
susceptibility $\bar{\chi}(\omega) .$ To show the definitions of  
$\bar{\epsilon}(\omega)$ and $\bar{\chi}(\omega),$ we have~\cite{Stroud} 
\begin{equation}
\langle{\bf D}\rangle = \bar{\epsilon}(\omega){\bf E}_0
 +\bar{\chi}(\omega)|{\bf E}_0|^2{\bf E}_0, 
\end{equation}%2
where $\langle\cdots\rangle$ denotes the spatial average, and 
${\bf E}_0 = E_0\hat{e}_z$ is the applied field along the $z-$axis.

We adopt the graded Drude dielectric profile
\begin{equation}
\epsilon(z,\omega) = 1 - {\omega_p^2(z)\over \omega(\omega+i\gamma(z))},\
 \ 0 \le z \le L.
\label{graded}
\end{equation}%3
In Eq.(\ref{graded}), we adopted various plasma-frequency gradation profile
\begin{equation}
\omega_p(z)=\omega_p(0)(1-C_{\omega} \cdot z/L),\label{Wp}
\end{equation}%4
and relaxation-rate gradation profile~\cite{Neeves}
\begin{equation}
\gamma(z)=\gamma(\infty)+\frac{C_{\gamma}}{z/L},\label{Rz}
\end{equation}%5
where $C_\omega$ is a dimensionless constant (gradient). Here 
$\gamma(\infty)$ denotes the damping coefficient in the corresponding 
bulk material. $C_\gamma$ is a constant (gradient) which is related to 
the Fermi velocity. Due to the simple layered geometry, we can use the 
equivalent capacitance of series combination to calculate the linear 
response, i.e., the optical absorption for the metallic film:
\begin{equation}
{1\over \bar{\epsilon}(\omega)} = {1\over L}
 \int_0^L {{\rm d}z\over \epsilon(z,\omega)}.
\end{equation}%6
The calculation of nonlinear optical response can proceed as follows.
We first calculate local electric field $E(z,\omega)$ by the identity
$$
\epsilon(z,\omega) E(z,\omega)=\bar{\epsilon}(\omega) E_0
$$
by virtue of the continuity of electric displacement, where $E_0$ is
the applied field.

In view of the existence of nonlinearity inside the graded film, 
the effective nonlinear response $\bar{\chi}(\omega)$ can be written 
as~\cite{Stroud} 
\begin{equation}
\bar{\chi}(\omega){\bf E}_0{}^4 = \langle\chi(z,\omega)
 |{\bf E}_{{\rm lin}}(z)|^2{\bf E}_{{\rm lin}}(z)^2 \rangle , 
\end{equation}%7
where $E_{{\rm lin}}$ is the linear local electric field.
Next, the effective nonlinear response can be written as an integral 
over the layer such as
\begin{equation}
\bar{\chi}(\omega) = {1\over L} \int_0^L {\rm d}z \chi(z,\omega)
 \left|{\bar{\epsilon}(\omega) \over \epsilon(z,\omega)}\right|^2
\left({\bar{\epsilon}(\omega) \over \epsilon(z,\omega)}\right)^2 .
\end{equation}%8

%\section{Numerical results}

For numerical calculations, we set $\chi(z,\omega)$ to be constant 
($\chi_1$), 
in an attempt to emphasize the enhancement of the optical nonlinearity. 
Without loss of generality, the layer width $L$ is taken to be $1$.

Figure~1 displays the optical absorption
$\sim\mathrm{Im}[\bar{\epsilon}(\omega)] ,$ the modulus of the
effective third-order optical nonlinearity enhancement
$|\bar{\chi}(\omega)|/\chi_1 ,$ as well as the FOM (figure of merit)
$|\bar{\chi}(\omega)|/\{\chi_1\mathrm{Im}[\bar{\epsilon}(\omega)]\}$
as a function of the incident angular frequency $\omega .$ 
Here $\rm{Im}[\cdots]$ means the imaginary part of $\cdots .$  
To one's interest, when the positional dependence of $\omega_p(z)$ is 
taken into account (namely, $C_{\omega}\ne 0$), a broad resonant plasmon 
band is observed. As expected, the broad band is caused to appear by the 
effect of the positional dependence of the plasma frequency of the 
graded metallic film. In particular, this band can be observed within 
almost the whole range of frequency, as the gradient $C_{\omega}$ is 
large enough. In other words, as long as the film under consideration is 
strongly inhomogeneous, a resonant plasmon band is expected to appear 
over the whole range of frequency. In addition, it is also shown that 
increasing $C_{\omega}$ causes the resonant bands to be red-shifted 
(namely, located at a lower frequency region). In a word, although the 
enhancement of the effective third-order optical nonlinearity is often 
accompanied with the appearance of the optical absorption, the FOM is 
still possible to be quite attractive due to the presence of the 
gradation of the metallic film.

Similarly, in Figure~2, we investigate the effect of the inhomogeneity 
of the relaxation rates $[\gamma (z)],$ which comes from the graded 
metallic film. It is evident to show that, in the low-frequency region, 
the positional dependence of relaxation rate $\gamma (z)$ enhances not 
only the third-order optical nonlinearity but also the FOM of such kind 
of graded metallic films.

Consequently, graded metallic films can be a suitable candidate material
for obtaining the {\it optimal} FOM. Thus, corresponding experiments 
are expected to be done to check our theoretical predictions since 
graded films can be fabricated easily.

%\section{Discussion and conclusion}

Here some comments are in order. We have discussed a graded metallic film 
(layered geometry), in an attempt to investigate the effect of gradation 
on the nonlinear enhancement and FOM (figure of merit) of such materials. 
It should be remarked that the optical response of the layered geometry 
depends on polarization of the incident light, because the incident 
optical field can always be resolved into two polarizations.
However, a large nonlinearity enhancement occurs only when the electric 
field is parallel to the direction of the gradient~\cite{Fischer}, and 
the other polarization does not give nonlinearity enhancement at 
all~\cite{Fischer}. 

In the conventional theory of surface plasmon resonant nonlinearity 
enhancement, there is often a dielectric component in the system of 
interest. 
In this regard, it turns out that it is not difficult to add a 
homogeneous dielectric layer on the metallic film. The same theory still 
works but a prominent surface plasmon resonant peak appears at somewhat 
lower frequencies in addition to the surface plasmon band. 
Due to the concomitantly strong absorption, the figure of merit of the
resonant enhancement peak is too small to be useful. In the limit of 
vanishing volume fraction of the dielectric component, however, the present 
results recover.

Since the surface plasmon frequency is 
proportional to the bulk plasmon frequency, one showed that the bulk plasmon frequency can be tuned by appropriate temperatures~\cite{PRL94,Chiang1}.  
That is, a position dependent plasma frequency can be achieved by imposing a 
temperature gradient in the direction perpendicular to the film. 
Nevertheless, the present results do not depend crucially on the 
particular form of the dielectric function. The only requirement is that we must have a 
sufficiently large gradient, either in $\omega_p(z)$ or in $\gamma(z)$ to 
yield a broad plasmon band.  We have shown that the dielectric function of a 
metal-dielectric composite with a variation in the volume fraction of 
metal along the z-axis can also give rise to a broad plasmon band. 

%In fact, a dielectric component can be introduced as 
%particulates embedded in the metallic component to form metal-dielectric 
%composites. The effective linear and nonlinear response can be 
%determined by appropriate mean-field theories as yet to be developed 
%because of the graded nature of the film.

It is instructive to extend our consideration to composites 
in which graded spherical particles are embedded in a host medium~\cite{HGYG} 
to account for mutual interactions among graded particles~\cite{Gao3}. 
In addition, it is also interesting to study the strong nonlinearity 
case~\cite{YuPLA96}.

To sum up, we have investigated the effective linear and third-order 
nonlinear susceptibility of graded metallic films with weak nonlinearity.
We calculated the effective linear dielectric constant and third-order 
nonlinear susceptibility.
It has been found that the presence of gradation in metallic films yields a 
broad resonant plasmon band in the optical region, resulting in a large 
nonlinearity enhancement and hence an optimal FOM.

{\it Acknowledgments}.
This work was supported by the Research Grants Council of the Hong Kong 
SAR Government under project number CUHK 403303, and by the DFG under 
Grant No. HO 1108/8-3 (J.P.H.).

\begin{figure}[h]
\caption{(a) The linear optical absorption 
${\rm Im}[\bar{\epsilon}(\omega)]$, (b) the enhancement of the
third-order optical nonlinearity $|\bar{\chi}(\omega)|/\chi_1$,
and (c) the FOM (figure of merit) $\equiv
|\bar{\chi}(\omega)|/\{{\chi_1\rm Im}[\bar{\epsilon}(\omega)]\}$
versus the normalized incident angular frequency
$\omega/\omega_p(0)$ for dielectric function gradation profile
$\epsilon(z,\omega)=1-\omega_p^2(z)/[\omega(\omega+i\gamma(z))]$
with various plasma-frequency gradation profile
$\omega_p(z)=\omega_p(0)(1-C_\omega \cdot z/L)$ and
relaxation-rate gradation profile
$\gamma(z)=\gamma(\infty)+\frac{C_\gamma}{z/L} .$ Parameters:
$\gamma(\infty)=0.02 \omega_p(0)$ and $C_\gamma=0.0$.}
\end{figure}

\begin{figure}[h]
\caption{Same as Fig.1. Parameters: $\gamma(\infty)=0.02
\omega_p(0)$ and $C_{\omega}=0.6$.}
\end{figure}

\newpage
\centerline{\epsfig{file=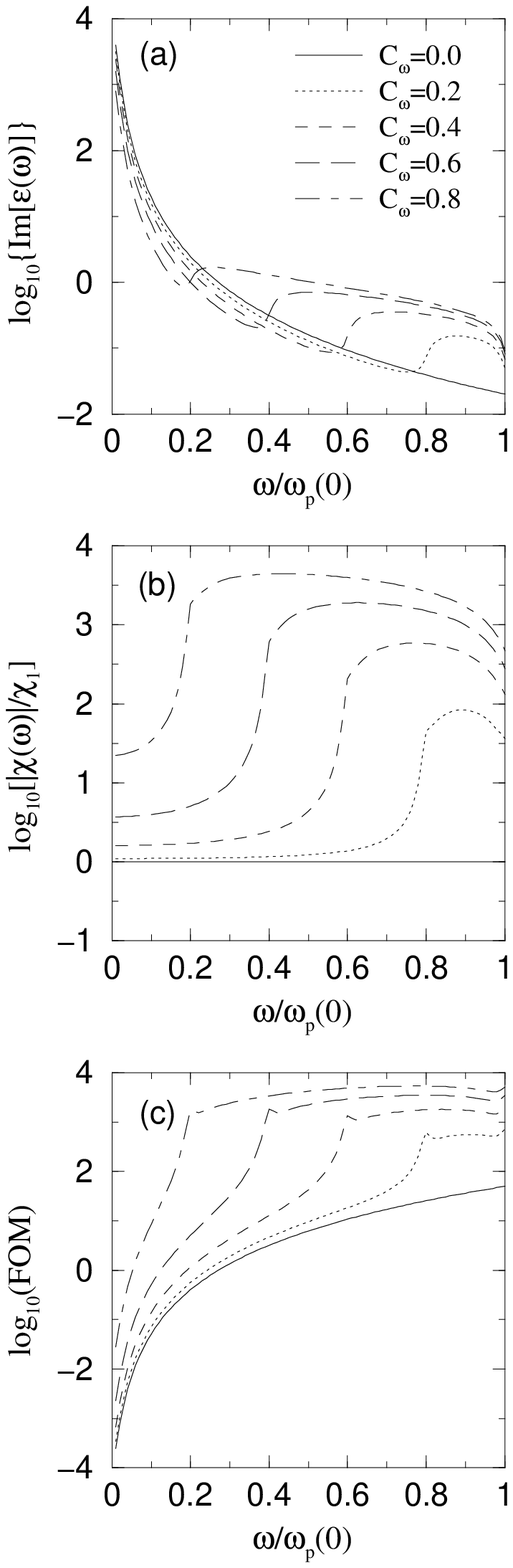,width=210pt}}
\centerline{Fig.1./Huang and Yu}

\newpage
\centerline{\epsfig{file=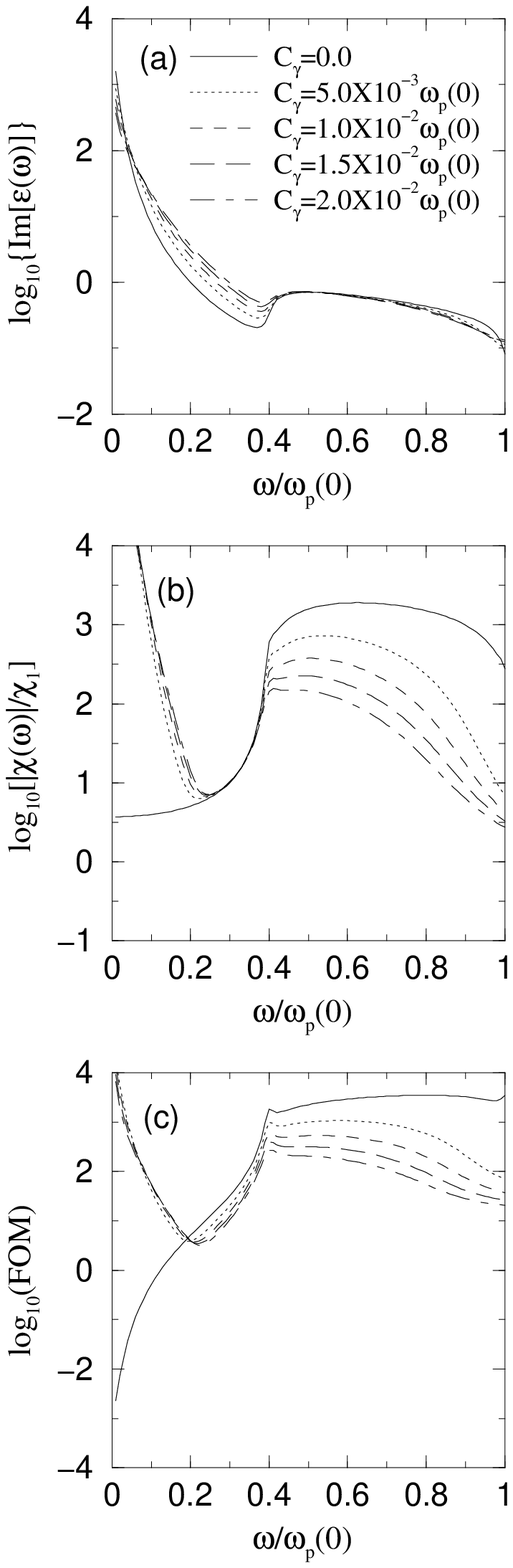,width=210pt}}
\centerline{Fig.2. /Huang and Yu}

\end{document}